# Blockchain Technology for Public Services: A Polycentric Governance Synthesis

Hozefa Lakadawala
Department of Electrical and Computer Engineering,
Binghamton University
hlakada1@binghamton.edu

Komla Dzigbede
Department of Public Administration and Policy
Binghamton University
dzigbede@binghamton.edu

Yu Chen
Department of Electrical and Computer Engineering
Binghamton University
ychen@binghamton.edu

**Abstract**
National governments are increasingly exploring blockchain technologies to improve transparency, trust, interoperability, and efficiency in public service delivery. Yet existing evidence on how blockchain is adopted and governed across national governments remains fragmented and often technologically focused. Using Polycentric Governance Theory as a conceptual lens, this study conducts a systematic review of peer-reviewed research published between 2021 and 2025 to examine public sector blockchain-enabled public services and the institutional, organizational, and information-management factors shaping their adoption. Following PRISMA guidelines, we synthesize findings from major digital government and information systems databases to identify key application domains, including digital identity, electronic voting, procurement, social services, and interoperable administrative systems, and to analyze the governance arrangements underpinning these initiatives. The analysis shows that blockchain adoption in public services is embedded within polycentric governance environments characterized by distributed authority, inter-organizational coordination, and layered accountability. Rather than converging on a single model of decentralization, governments configure blockchain systems through hybrid and permissioned designs that enable selective decentralization while retaining centralized oversight, a pattern we conceptualize as controlled polycentricity. By reframing blockchain as a governance infrastructure that encodes coordination, accountability, and information-sharing rules, this study advances polycentric governance and digital government theory beyond descriptive accounts of technology adoption. The findings offer theoretically grounded insights for researchers and practical guidance for policymakers and public managers seeking to design, govern, and scale blockchain-enabled public services.

# 1. Introduction

Blockchain is a distributed ledger technology in which transactions are recorded in continuously increasing data blocks linked through cryptographic signatures or hashes. Blockchains serve as the foundation upon which smart contracts – self-executing code with predefined rules – operate (IBM, 2025). Decentralized applications further extend blockchain functionality by interacting with smart contracts to deliver open-source, user-facing services. Examples of blockchain-based technologies include Bitcoin and cryptocurrency (Nakamoto 2008), Ethereum (Wood 2014), and Hyperledger Fabric (Androulaki et al. 2018). Over the years, blockchain has evolved into a novel approach for storing and processing information in a decentralized, secure, and efficient manner (Calvin et al., 2020).

National governments are increasingly exploring blockchain technologies to advance e-government (EGOV) priorities. EGOV refers to the use of information and communication technologies (ICTs) to improve public service delivery, strengthen transparency, enhance citizen engagement, and support more accountable and effective government operations. In this context, blockchain can serve as a trusted record-keeping and public management tool that facilitates secure interactions with citizens and businesses, reduces administrative burdens, and supports transparent collaboration among diverse stakeholders (Krichen et al., 2022).

Despite growing interest, the adoption of blockchain in government occurs within complex institutional environments involving multiple agencies, overlapping jurisdictions, and distributed decision-making authority. Polycentric Governance Theory provides a useful conceptual lens for understanding how blockchain adoption decisions are shaped within such multi-actor settings. A polycentric governance perspective emphasizes the roles of multiple, semi-autonomous decision centers, shared rules, accountability mechanisms, and coordinated information-sharing, all of which influence whether, how, and to what extent national governments adopt blockchain technologies (Ostrom, 2010; Aligica & Tarko, 2012). Integrating this theoretical lens enables a deeper examination of how institutional arrangements, administrative capacity, and inter-organizational coordination shape blockchain adoption across national governments.

This article uses a polycentric governance lens to guide a systematic review of blockchain adoption in the public sector. It analyzes peer-reviewed research published from 2021 to 2025 to identify emerging blockchain applications within national governments and to assess the challenges and opportunities associated with its adoption in public services. Specifically, the study addresses two interrelated research questions: (1) What decentralized blockchain applications are national governments adopting for public service delivery? (2) What strengths, weaknesses, opportunities, and threats influence blockchain adoption in the public sector?

Following PRISMA (Preferred Reporting Items for Systematic Reviews and Meta-Analyses) guidelines, we systematically reviewed peer-reviewed publications from primary sources, including IEEE Xplore, SpringerLink, and ACM Digital Library. The findings reveal that governments are adopting blockchain-based applications across a variety of service domains, including digital identity, electronic voting, e-tendering, social security, public spending, and interoperable administrative systems. The review also highlights the institutional, organizational, and governance challenges that impede blockchain adoption and the opportunities to strengthen transparency, coordination, and public value creation.

The remainder of this article is structured as follows. Section 2 presents the theoretical foundation, and Section 3 describes the methodology for the systematic review of the extant literature. Section 4 reports the review findings by examining trends in blockchain adoption and synthesizing patterns across diverse public service domains. Section 5 articulates how blockchain-enabled systems reconfigure governance roles, rules, and responsibilities in digital government settings. Section 6 discusses the challenges and opportunities related to blockchain adoption in national governments, and Section 7 concludes with key implications for research and practice.

## 2. Polycentric Governance Theory

Governance challenges in the adoption of blockchain-based public services arise not only from technological factors but also from the complex institutional arrangements that characterize contemporary public management. To analyze these dynamics, this study draws on Polycentric Governance Theory, developed by Ostrom (1999, 2010), to explain how *multiple centers of decision-making authority* function concurrently and coordinate through shared rules and governance arrangements. In polycentric systems, autonomous yet interdependent actors function under overlapping jurisdictions and engage in continual coordination, negotiation, learning, and adaptation (McGinnis, 2011). This framework is well-suited to examining blockchain adoption in national governments, where responsibilities for digital infrastructure, data governance, cybersecurity, and public service delivery are dispersed across diverse agencies and stakeholders.

In addition, blockchain's distributed ledger architecture aligns closely with polycentric governance principles. Rather than relying on a single centralized authority, blockchain systems use *decentralization*, consensus protocols, and shared validation to maintain integrity and ensure reliable recordkeeping (De Filippi & Wright, 2018). This design parallels the multi-actor institutional configurations that influence the adoption and governance of digital public services. Applications such as digital identity systems, electronic voting, procurement platforms, and social service delivery involve coordination among government ministries, regulatory agencies, audit institutions, private firms, and citizens. A polycentric perspective highlights how governance in these settings is shaped by decentralized institutional arrangements and by the distributed architecture of blockchain systems.

Furthermore, blockchain adoption by national governments often spans *multiple policy domains and administrative levels*. Introducing decentralized applications in areas such as identity management or public procurement requires alignment of rules, standards, and governance arrangements among ministries of the interior, ICT authorities, procurement agencies, and auditing bodies. Polycentric governance emphasizes that overlapping authority structures necessitate shared norms and coordinated rule systems to avoid fragmentation and ensure transparency and effectiveness (Ostrom, Tiebout, & Warren, 1961). Applying this framework, therefore, deepens understanding of the institutional interdependencies that shape blockchain adoption in the public sector.

Polycentric governance also underscores the value of *experimentation and learning*. Governments commonly initiate blockchain projects through pilots or limited-scope implementation, providing opportunities to test approaches and adapt designs to local contexts. Such decentralized experimentation is a defining feature of polycentric systems, which foster innovation by enabling multiple centers of action to generate and refine solutions (Ostrom, 2010; McGinnis & Ostrom, 2014). Using this lens could expand understanding of why countries pursue different blockchain adoption strategies and of the mechanisms through which lessons circulate across institutional contexts.

*Accountability and trust* are key attributes that further highlight the alignment between blockchain and polycentric governance. Blockchain's transparent, immutable, and auditable recordkeeping can strengthen accountability in public services. Polycentric governance similarly emphasizes that accountability in distributed systems arises from shared norms, collective rule-making, and peer monitoring rather than hierarchical oversight alone (Aligica & Tarko, 2014). Blockchain's consensus mechanisms reflect these principles by enabling multiple actors to participate in validation and oversight. At the same time, a polycentric lens highlights challenges – such as ambiguous role assignments or inconsistent rule enforcement – that may emerge when authority is distributed broadly.

Taken together, Polycentric Governance Theory provides a coherent conceptual foundation for synthesizing research on blockchain adoption in government. It shows how blockchain reshapes institutional arrangements, inter-agency coordination, accountability structures, and opportunities for decentralized innovation, while also emphasizing the shared rules and standards needed for long-term viability. Figure 1 illustrates the key constructs of polycentric governance and their application to blockchain adoption and oversight in the public sector.

**Figure 1.**
*Polycentric Governance and Blockchain-Enabled Public Services.*

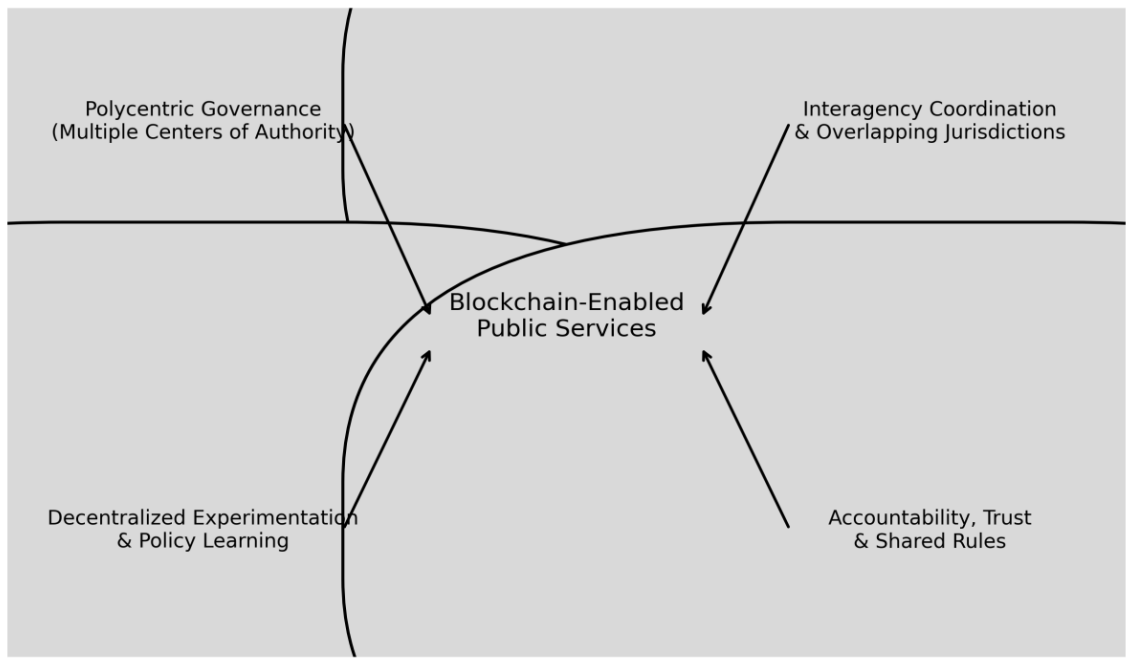

*Note.* Authors' conceptual model illustrating how key dimensions of polycentric governance may inform blockchain-enabled public service delivery.

### 3. Method

The systematic literature review protocol was developed in accordance with PRISMA recommendations to ensure methodological rigor. The literature search was conducted in three phases to capture both earlier and more recent studies. The initial search conducted in December 2023 used Google Scholar as the primary search engine due to its extensive coverage of disciplines and reputable academic publications, including IEEE Explore, Springer,

ScienceDirect, Taylor & Francis online, and ACM Digital Library. The query for the search included simple keywords such as 'blockchain' and 'public services'. Only English language peer-reviewed publications were selected. In addition, we used the filter 'since 2021' on Google Scholar to limit the results to publications from 2021 and onwards, reflecting our focus on recent empirical advancements. A follow-up search on October 1, 2024, used the same strategy to collect the latest scholarly developments and maintain the review's currency. Finally, we conducted a search in July 2025 to obtain the most recent information and enhance the study's currency.

Figure 2 illustrates our paper selection process. A total of 276 records were identified across the searches conducted: IEEE (n=76), Springer (n=45), MDPI (n=19), ScienceDirect (n=59), and other publication sources (n=77). After removing 51 duplicates, 225 unique records were subjected to title and abstract screening against four inclusion and four exclusion criteria detailed in Table 1. On one hand, the inclusion criteria ensured that the studies were (1) readable in English, (2) peer-reviewed, (3) published post-2020, and (4) contained a conclusion section. On the other hand, the exclusion criteria eliminated (1) non-academic sources, (2) grey literature, (3) irrelevant topics, and (4) studies lacking a focus on the public sector or national governments.

**Figure 2**
*PRISMA Flow Diagram of the Systematic Review Process*

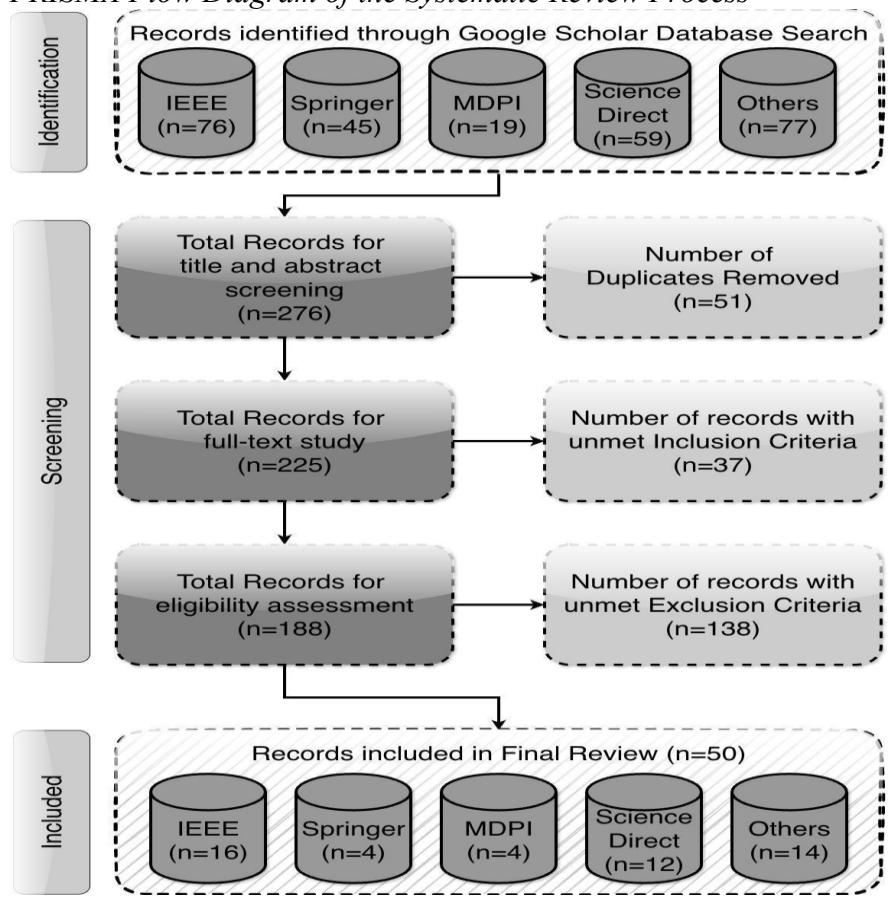

*Note*. This figure illustrates the present study's search and screening procedures based on PRISMA guidelines.

**Table 1**

*Eligibility Criteria and Justification for Study Selection*

| Eligibility Criteria | Justification |
|---|---|
| **IC1: Articles must be published in the English Language** | To facilitate accurate assessment of study quality and findings, along with consistency in language |
| **IC2: Peer-reviewed articles from conferences, journals, and book chapters** | To guarantee that only rigorously reviewed and validated studies are included in the review |
| **IC3: Articles must have a conclusion section** | To include only studies that provide a clear summary of findings |
| **IC4: Publishing year must be 2021 or later** | To focus on recent studies reflecting current practice and knowledge |
| **EC1: Non-academic and unpublished articles** | To remove sources lacking rigor, authority, and credibility |
| **EC2: Articles not related to blockchain** | To keep the focal point of selection on blockchain |
| **EC3: Articles not discussing public service provisioning or decision making** | To exclude articles whose primary theme of study is not public services |
| **EC4: The level of discussion must be at the national, international, or supra-national government level** | To keep out articles where the unit of analysis is the subnational government |

*Note.* This table summarizes the criteria used to include or exclude studies during the study selection process. IC stands for Inclusion Criteria, and EC means Exclusion Criteria.

Overall, 37 articles were excluded for failing to meet the inclusion criteria. This left 188 records for eligibility evaluation, of which 138 were excluded for reasons such as insufficient methodological detail or lack of relevance to public sector or national government contexts. Ultimately, 50 studies met all criteria and were included in the final review: IEEE (n=16), Springer (n=4), MDPI (n=4), ScienceDirect (n=12), and others (n=14). Two reviewers independently conducted screening and eligibility assessments, and disagreements were resolved through discussion to minimize bias. This process ensured that only relevant high-quality studies were included in the systematic literature review. Furthermore, data extraction was performed systematically using a predefined standardized form to ensure consistency and adhere to PRISMA 2020 guidelines. We manually captured key characteristics of the identified articles, including publication year, author affiliation, publication source, and methodology.

## 4. Findings

While the reviewed studies span diverse public service domains and technological architectures, a cross-cutting pattern emerges across cases: blockchain adoption in national governments is consistently shaped by governance considerations related to authority distribution, inter-organizational coordination, accountability, and information control. Accordingly, the findings below are presented by application domain, but interpreted through a polycentric governance lens that emphasizes how institutional arrangements condition blockchain design and implementation in public services.

### 4.1 Descriptive Characteristics and Thematic Patterns

In the present study, we synthesize research across major publication outlets, geographical distribution of authors, methodological approaches, and the key themes emerging from the systematic review. Figure 3 shows the geographical diversity assessed across the total number of identified records mapped by the first author's location. Europe-based (25 studies) and Asia-based (17 studies) authors dominated our findings, reflecting vigorous research activity in these regions, and indicating significant government investment in developing and adopting blockchain technology for public services in these regions. North American authors contributed 4 studies, while Africa, Oceania, and South America provided minimal representation (2, 1, and 1 studies, respectively). This distribution highlights a potential geographical bias, with underrepresented regions potentially facing barriers to blockchain adoption (e.g., limited funding or missing research infrastructure).

**Figure 3**
*Distribution of Studies by First Author's Geographical Location (2021 to 2025)*

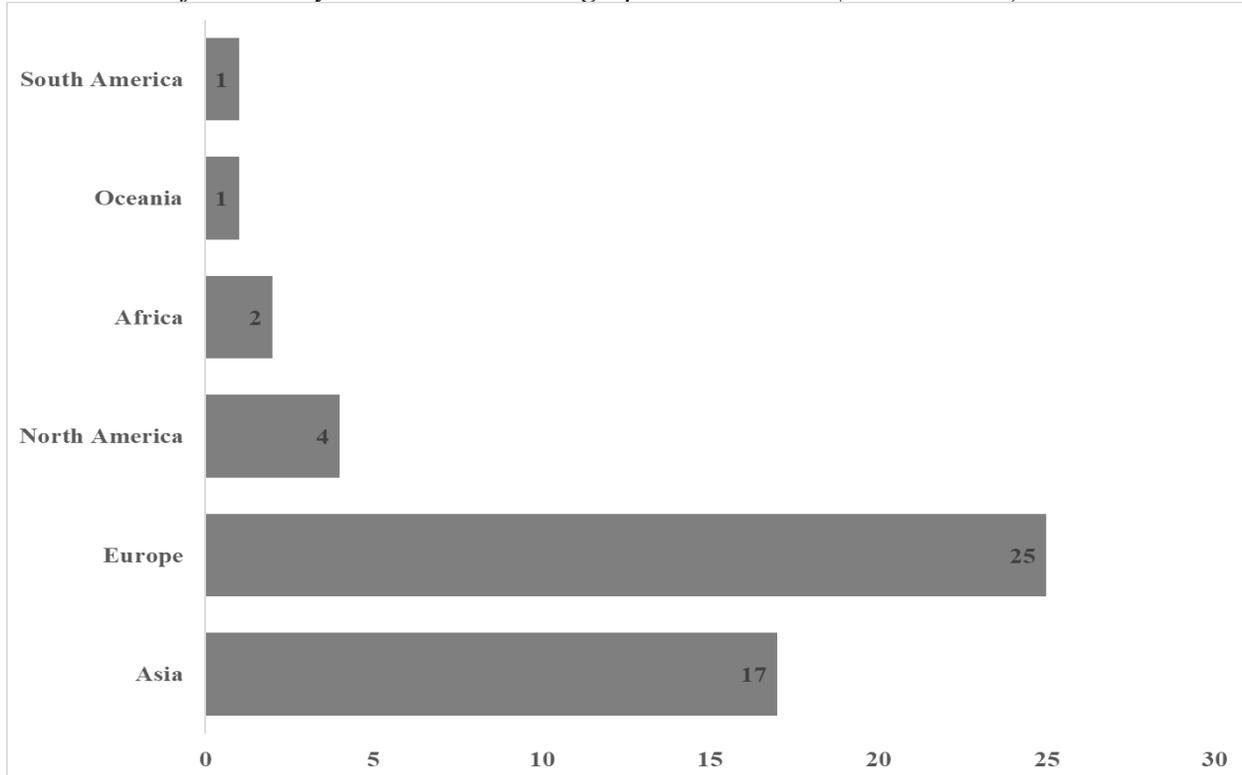

*Note*. Geographical locations are derived from the first author's institutional affiliations in the included studies.

Figure 4 shows the distribution of studies across the major publishers and years (2021–2025). IEEE contributed the largest share (16 studies), peaking in 2022. ScienceDirect (12 studies) holds the second largest share of studies. In comparison, Springer (4 studies), MDPI (4 studies), ACM (3 studies), and Taylor and Francis (2 studies) show fewer but significant contributions, indicating a consistent interest in the topic. Furthermore, Figure 5 reveals that 34% of the studies analyzed blockchain technology as a novel and innovative application in public service domains. 27% presented empirical use cases in the public sector, evaluating real-world applications with the participation of stakeholders, including civil servants, citizens, and IT experts. 23% of the studies focused on theoretical and conceptual frameworks for blockchain adoption in the public sector. In contrast, the remaining 16% were systematic reviews of the literature that synthesized previous research.

The systematic analysis in this study reveals four major themes as illustrated in Figure 6. The themes reflect distinct scholarly contributions to the evolution and application of blockchain technology in public services. The first theme offers analytical perspectives on the adoption of blockchain in public services, including e-government and smart public services. The second theme focuses on the implementation of state-of-the-art innovative technologies in public services. The third theme examines empirical public service use cases, providing qualitative insights from civil servants and IT experts, as well as quantitative evaluations based on citizens' perceptions of blockchain in public services. Finally, the fourth theme consists of assessments of technological, organizational, environmental, socioeconomic, political, and legal challenges related to blockchain governance in the public sector.

**Figure 4**
*Distribution of Studies Across Major Publishers and Years (2021–2025).*

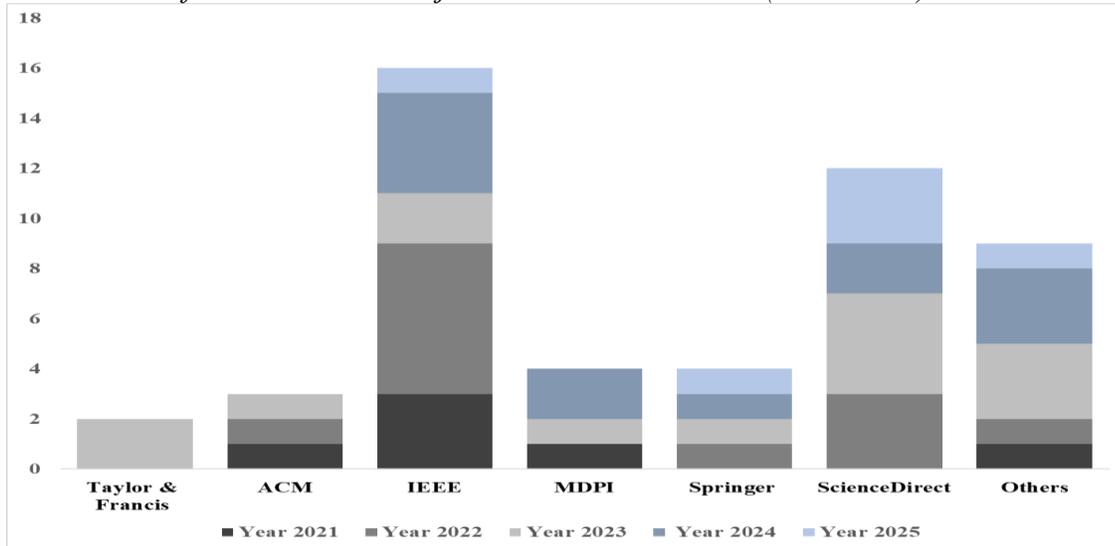

*Note.* The figure summarizes the distribution of studies by publisher and year from 2021 to 2024.

**Figure 5**
*Key Methods and Approaches in the Studies Analyzed*

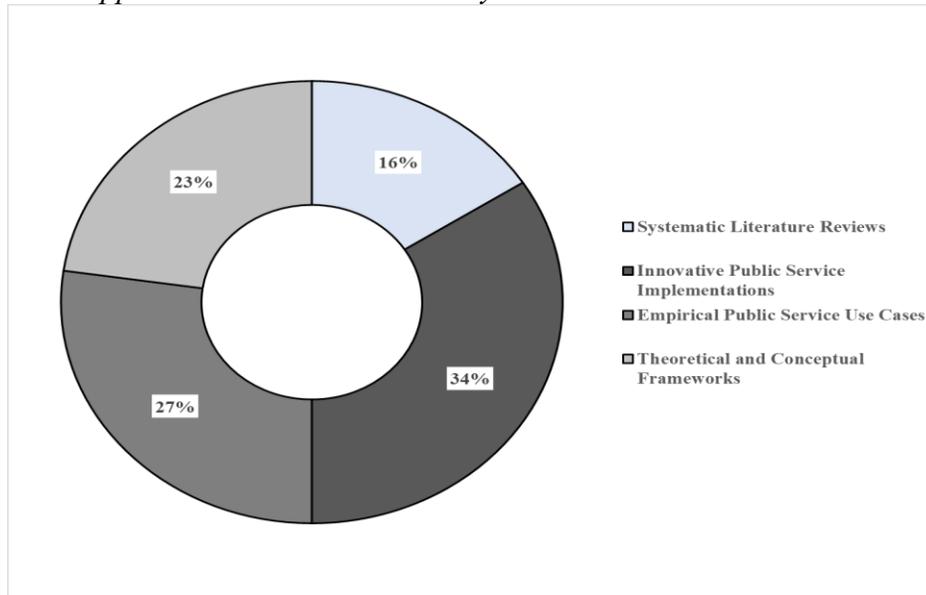

*Note*. The figure categorizes studies by their key methods and approaches, including innovative applications, empirical case studies, theoretical and conceptual frameworks, and systematic reviews.

**Figure 6**
*Major Themes and Emphases in the Identified Records*

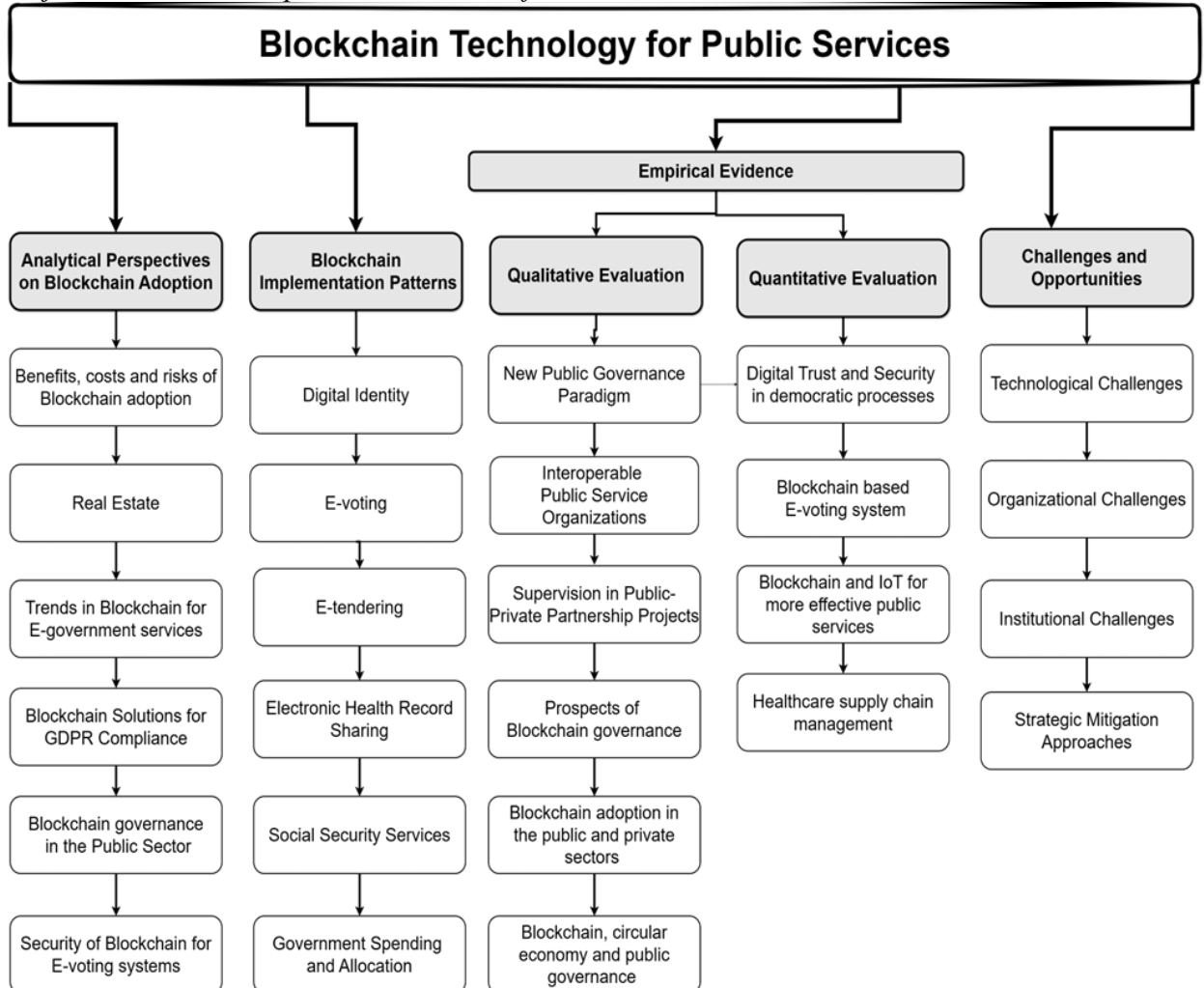

*Note.* The figure illustrates four major themes and emphases identified in the systematic review: analytical perspectives on blockchain adoption in the public sector, patterns of technology implementation in public services, empirical use cases, and assessments of blockchain governance challenges and opportunities.

Interpreted through a polycentric governance lens, these descriptive characteristics and thematic patterns suggest that blockchain adoption in national governments is shaped less by technological diffusion alone and more by underlying governance arrangements. The prevalence of diverse application domains, stakeholder configurations, and methodological approaches reflects the presence of multiple, semi-autonomous decision-making centers operating under shared institutional rules. Rather than converging on a single dominant model, blockchain-enabled public services exhibit variation consistent with polycentric governance systems, in which experimentation, learning, and adaptation occur across overlapping jurisdictions and organizational boundaries.

## 4.2 Blockchain Implementation Patterns in the Public Sector

This section addresses the research question concerning the state-of-the-art decentralized blockchain applications currently implemented in public service delivery across countries. To synthesize these implementation practices, we present a unified public service architecture that captures common blockchain implementation patterns observed in national government contexts, as shown in Figure 7.

At the top of the architecture, the application presentation layer enables user access through mobile applications, mini programs, and web browsers. This layer supports seamless interaction with a range of government services, including secure electronic health record sharing, real estate management, social security administration, digital identity management, and government spending and allocation systems. By abstracting underlying blockchain complexity, the presentation layer facilitates user-friendly engagement with blockchain-enabled public services.

Beneath this layer, the architecture integrates Ethereum-based smart contracts with the Inter Planetary File System (IPFS) to support decentralized application logic and off-chain data storage. Smart contracts govern transactional processes, while IPFS provides a distributed storage solution for large or sensitive files, such as identity records, real estate documentation, and social security data. This separation of computation and storage enhances scalability while preserving decentralization and data integrity.

**Figure 7**
*Blockchain Implementation Architecture for National Public Services*

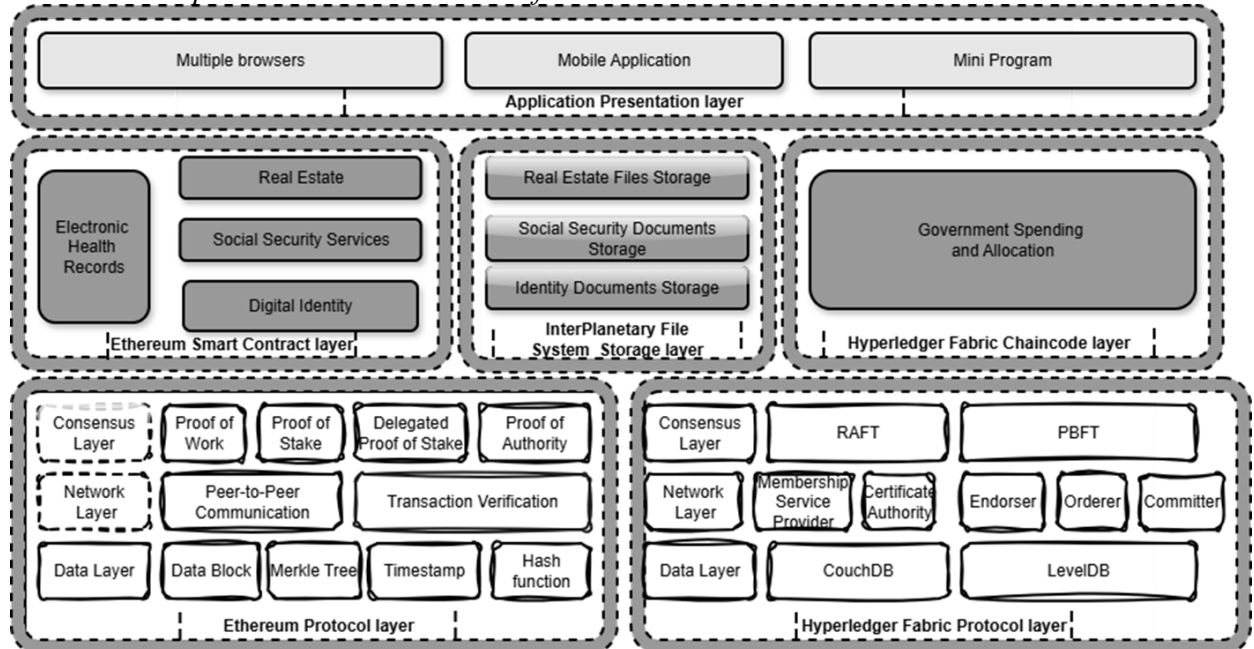

*Note.* The figure synthesizes common blockchain implementation patterns observed across national public service systems.

The architecture adopts a hybrid blockchain model that combines permissionless and permissioned infrastructures to accommodate diverse public sector requirements. Ethereum smart contracts support public-facing, open-access operations, whereas Hyperledger Fabric chain code underpins permissioned, enterprise-grade processes that require stricter access controls.

This hybrid design enables national governments to balance transparency and openness with privacy, regulatory compliance, and institutional accountability.

At the protocol level, the Ethereum platform relies on consensus mechanisms such as Proof-of-Work, Proof-of-Stake, Proof-of-Authority, and Delegated Proof-of-Stake to validate transactions and secure the network. Peer-to-peer communication and transaction propagation are facilitated by the network layer, while data integrity is maintained through core data layer components, including Merkle trees, cryptographic hash functions, timestamps, and data blocks.

Complementing this, the Hyperledger Fabric protocol layer employs consensus mechanisms such as RAFT and Practical Byzantine Fault Tolerance (PBFT), supported by network entities including certificate authorities, membership service providers, orderers, endorsers, and committers. Data persistence is managed through databases such as CouchDB and LevelDB, enabling scalable and privacy-preserving operations for sensitive interorganizational processes within government systems.

The remainder of this section analyzes the diverse domains within which governments can use (or are currently using) blockchain technology for public service delivery. Our systematic review of the literature identified eight key domains, namely: (1) digital identity solutions, (2) decentralized electronic voting, (3) decentralized electronic tendering, (4) secure electronic health record sharing, (5) social security services, (6) government spending and allocation, (7) real estate registrations, and (8) interoperable public service systems.

*4.2.1 Digital Identity solutions*

Safeguarding the digital identity of citizens is one of the most fundamental public services governments must provide. In traditional systems, personal identification platforms collect and maintain sensitive information in a trusted central intermediary database to legally recognize the identity of citizens. However, the lack of adequate security in such systems can lead to identity theft due to data breaches. On the other hand, self-sovereign identity with a blockchain-based identity management system offers user-centric control and access to identification along with secure and fraud-resistant solutions to identity theft (Skelaney et al. 2023). Furthermore, distributed systems can be used to combat the lack of proper identification threats, providing a secure, resilient, and authenticity-guaranteed platform. Thumb fingerprints can be collected to generate a readable string character in the field, which would be encrypted and stored on the blockchain (Habib, Refat & Ahad, 2023).

In South Africa, for example, there is a strong demand to implement decentralized systems with proper third-party access controls and leverage the key blockchain property of data immutability. The national government has utilized the *Ethereum Goerli testnet* to aid sensitive document verification, like a driver's license. In this case, the data storage objective is fulfilled using the interplanetary file system (IPFS), and the generated hashes are attested to the blockchain using smart contracts (Mahlaba et al., 2022). Another example is Indonesia, where the national government uses a private Ethereum blockchain network to address major risk factors in its database system. Smart contracts automate the workflow, storing information on the national identity card and land certificates, and managing data using a non-fungible token (NFT) document inspector application (Azhar et al., 2023).

*4.2.2 Decentralized Electronic Voting*

Governments can use blockchain-based electronic voting systems to minimize vote rigging, reduce budget expenditures in national elections, and promote user privacy and enhance security (Widayanti et al., 2021). Conventional general elections result in frequent errors and incoming data manipulation. While e-voting in these conventional systems can be easy to use, it is heavily centralized, prone to error, and vulnerable to outside attack. However, governments can explore state-of-the-art blockchain-based electronic voting systems where votes are stored in blocks along with the previous hash, timestamp, voter's signature, and ID using a secure hash algorithm (SHA-256) and elliptic curve cryptography (ECC) (Widayanti et al., 2021; Ohize et al., 2025).

### *4.2.3 Decentralized Electronic Tendering*

Governments often procure goods and services from external suppliers to support service delivery to citizens. As part of the procurement process, external suppliers submit bids or tenders from which governments use established laws and procedures to select the best bid. E-tendering has emerged as a digital method of carrying out the tendering phase of government procurement using electronic platforms. These online portals enable governments to publish tenders online, receive electronic submission of tenders, undertake online evaluation of tenders, and digitally award contracts to external suppliers. However, traditional centralized E-tendering procedures face information asymmetry and security problems, such as discretionary information disclosure and data breaches or tampering.

With a blockchain-based decentralized system, government e-tendering processes can be made more secure. Tender documents can be stored in an encrypted database with the necessary information stored in a smart contract. Registered bidders using the Advanced Encryption Algorithm (AES) can access the available list of tenders, including details on the platform. Once selected, they can place a bid by providing an encrypted quote along with their clauses. The bidder can also provide a balance sheet and a registered license to establish credibility during the allocated bidding period. Once the deadline is reached, the government gains access to all submitted bids that the provided bidder's key can decrypt and selects the most competitive bid as the winner. The decentralized blockchain system would prevent bidders from tampering with each other's bids, disable the bidders from modifying tender details once posted, and encrypt bids using the computationally efficient Advanced Encryption Standard (AES) algorithm suitable for real-time processing of large data volumes (Marati et al., 2023).

### *4.2.4 Secure Electronic Health Record Sharing*

China has implemented a tiered healthcare system to optimize the use of medical resources in primary healthcare facilities and ensure accessible basic healthcare to patients. Built on a blockchain platform, the system comprises seven categories of participants: medical facilities, medical staff, patients, health insurance institutions, regulatory authorities, manufacturers, and quality inspection departments. A consortium chain with a proof-of-authority (PoA) consensus algorithm is strategically selected to reduce the financial burden on the government, harmonizing throughput, data security, and scalability for medical data sharing. In this system, two categories of smart contracts are defined for participant identity management and the authoritative participant selection process, providing a timely response to patients' healthcare needs and preserving the confidentiality of records, respectively (Geng, Chuai & Jin, 2024).

Similarly, European Union countries have introduced a blockchain-based framework for the secure and privacy-preserving exchange of electronic health records (EHRs) among healthcare providers, citizens, and government departments and agencies. The system utilizes a public Ethereum blockchain for record indexing and an EU-compliant digital identity system (eIDAS) for access control. A cryptographic hashing scheme is employed to decouple patient identities from their electronic health records on the blockchain, ensuring unlinkability. Access to the health records is granted only to entities authorized by the patient through verifiable assertions (Lax, Nardone & Russo, 2024).

### *4.2.5 Social Security Services*

Social security services play a crucial role in public administration, as citizen-centric services help address the common interests and needs of citizens. In the United States, the Social Security Administration provides many of its services online, including filing appeals, applying for Medicare, retirement, and disability benefits, and requesting assistance with housing, nutrition, and low-income home energy. Traditional social security information systems face challenges such as inconsistent interfaces and cumbersome authority authentication mechanisms, which ultimately result in incomplete data sharing. Additionally, they are unable to perform manual reviews in the background, leading to delayed approvals and a lack of authentic verification of proof materials.

A consortium blockchain using Hyperledger Fabric (HLF) can be introduced to build a unified and shared social security service data platform, retaining tamper-proof, traceability, and consensus-based trust characteristics with an interplanetary file system (IPFS) to overcome storage limitations. Researchers have illustrated the utility of state-of-the-art blockchain applications for online social security services. For example, five smart contracts written in the Go language have been used to implement specific online functions of social security services. Registration smart contracts ensure that only authorized participants can communicate on-chain and execute function calls, and they also map national databases to the peer node (Tang et al. 2022).

### *4.2.6 Government Budgetary Allocation and Spending*

Blockchain can help eliminate fraudulent behaviors, such as theft, bribery, false claims, and embezzlement, ensuring that government funds are spent as intended and accountability is maintained. The permissioned blockchain using HLF is designed to be well suited to scenarios in which different access control levels need to be created for writing permissions, allowing only authorized and selected nodes to access them. The logs in the multi-ledger chain originate from the national government's allocation as the root node, which is then distributed among different audited ministries as child nodes using various channels. Financial transactions, such as spending and the allocation of government-endowed grants, can be enhanced through full traceability and accountability while also providing robust security (Ranjan et al., 2022).

Evidence from Nigeria illustrates how blockchain has been explored to address specific government allocation challenges. Bello and Thomas (2023) have proposed an HLF-based implementation of a consortium chain comprising two distinct groups of participants–actors are agencies such as the central bank and the Federal Character Commission, while observers are the national police force, government ministries, the judiciary, and the economic and financial crimes commission. The actors would serve as orderers using the practical Byzantine fault tolerance (pBFT) consensus protocol. In contrast, the observers serve as non-ordering peers in the proposed HLF-based implementation (Bello & Thomas, 2023). Such a blockchain-based system would reduce problems with ghost workers and inflated wage bills of the government.

### *4.2.7 Real Estate Registrations*

Some countries have proposed collaborative blockchain projects to improve the management of real estate registration and transactions. In Kazakhstan, researchers have proposed a system in which the State Corporation, the Ministry of Justice, and one of the private banks would utilize a blockchain-based system to reduce the time for real estate pledge registration to a single day, with reduced costs, exclusion of paperwork, reliable, secure data, and minimal civil servant interaction (Akhmetbek & Spacek, 2021).

Similarly, in Australia, Zhang et al. (2024) have proposed that the government address flaws in the existing real estate management system by adopting the Ethereum blockchain, integrating IPFS, and using *Infura* for distributed data storage services. The key participants in this proposed real estate management system would include the land and real estate registration department, financial institutions, sellers, and buyers of real estate. The land management department, as the authorized entity, would be responsible for establishing user information and real estate information databases. Using smart contracts, real estate transactions would be securely automated using *OpenZeppelin* libraries, storing hash values of related IPFS uploaded real estate documents and photos with timestamps, and using *MetaMask* virtual currency wallet, maintaining unforgeable transactions (Zhang et al., 2024).

### *4.2.8 Interoperable Public Service Systems*

Unlike traditional centralized data exchange systems, such as point-to-point integration and the hub-and-spoke approach, Blockchain offers new opportunities for public service organizations to communicate and integrate their operations to improve efficiency. Shahaab et al. (2021) illustrated this potential through a case study of UK public service organizations. As the official business trading registrar in the UK, Companies House maintains approximately 4.8 million business records vital to the UK economy. It also shares private information with different government departments and predetermined organizations, rather slowly and inefficiently, through data dumps or manual request handling. Overly bureaucratic and expensive information sharing leaves an inconsistent trail among public service organizations, necessitating the design of connected public services to detect anomalies or suspicious activities.

Against this background, Shahaab et al. (2021) developed a proof-of-concept that used a public-private hybrid blockchain architecture, in which *Corda* enables private, intergovernmental data sharing in an automated, real-time manner, and Ethereum is employed for immutability, data integrity, and public verification. The interoperable consortium comprises Companies House, the Revenue and Customs authority (HMRC), the national land registry, and notary society nodes, which act as remote procedure call (RPC) clients on *Corda*. Those nodes are seamlessly integrated while sharing information on a need-to-know basis, and they flexibly map to native data management systems. Users would submit details only once to any of the public registries, within a continuous feedback-loop system that would enable state synchronization and further mitigate the costs associated with identifying discrepancies. IPFS will be used as off-chain storage of personal and transactional data, while the commitment will be added to the public Ethereum blockchain (Shahaab et al., 2021).

From a governance perspective, the implementation patterns identified in this subsection illustrate how blockchain functions as an institutional coordination infrastructure within polycentric public service systems. The widespread use of hybrid architectures, permissioned networks, and role-based access controls indicates a deliberate allocation of authority among governmental and non-governmental actors. These design choices enable decentralized execution and information sharing while preserving centralized accountability and regulatory oversight. As such, blockchain implementation in the public sector reflects a form of controlled polycentricity in which technological decentralization is conditioned by institutional and legal constraints.

**4.3 Empirical Evidence on Blockchain Adoption, Governance, and Performance**

This section synthesizes empirical evidence on blockchain adoption, governance, and performance in the public sector. Researchers have employed both qualitative and quantitative methods to examine how blockchain technologies are adopted and governed across public service contexts. Our systematic review shows that common qualitative approaches include interviews, focus groups, and design science research, often used to explore blockchain governance paradigms, institutional arrangements, and system design choices. By contrast, quantitative studies have relied on survey-based analyses, technology acceptance models, and structural equation modeling to assess factors influencing adoption, citizen perceptions, and performance indicators. Overall, empirical evidence from the extant literature highlights recurring adoption factors, governance considerations, and performance dimensions.

*4.3.1 Blockchain Adoption Factors*

Figure 8 summarizes insights from selected case studies in which researchers used quantitative methods to examine factors influencing blockchain adoption in the public service domain. For example, Mannonov and Myeong (2024) analyzed a case from Uzbekistan using survey data from 387 respondents and applied a technology acceptance model to assess citizens' perceptions of blockchain-based electronic voting. Their findings indicate that *perceived usefulness* and *ease of use* were key factors shaping government intentions to adopt blockchain-enabled voting systems. Across other quantitative case studies, additional determinants of blockchain adoption include perceived security, trust, attitudes toward use, and societal involvement, as well as broader technological factors (e.g., system interoperability), organizational considerations (e.g., organizational readiness), and environmental conditions (e.g., regulatory context).

**Figure 8**
*Synthesis of the Determinants of Blockchain Adoption in the Public Sector.*

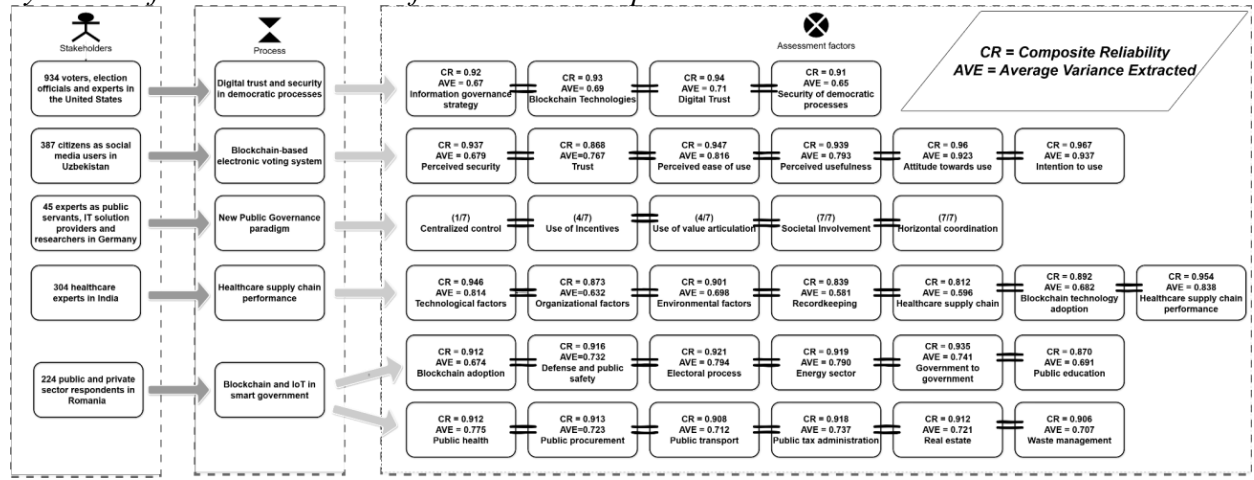

*Note.* This figure presents a synthesis of the determinants of blockchain adoption in the public sector, drawing on evidence from extant empirical studies. The determinants reflect technological, organizational, and environmental dimensions discussed in the reviewed literature.

*4.3.2 Governance Dimensions of Blockchain Adoption*

Empirical studies situate blockchain governance in the public sector within the broader evolution of public administration toward the New Public Governance paradigm. This paradigm shifts attention beyond efficiency and effectiveness to encompass values such as transparency, resilience, sustainability, and the public interest (Osborne & Gaebler, 1992; Denhardt & Denhardt, 2015; Nukpezah et al., 2021). Consistent with this perspective, interviews of representatives of blockchain pilot projects in Malta, Sweden, and Switzerland suggest that blockchain-based public service delivery systems can be governed in ways that align with New Public Governance expectations (Brinkmann, 2021).

Building on this governance framing, citizen-, expert-, and developer-based studies further clarify how blockchain governance operates in practice. A Delphi survey of blockchain experts finds broad agreement that public administrators play a central role as trusted coordinators in blockchain governance, with horizontal coordination and social participation viewed as particularly important, while incentives and value articulation are perceived as less central (Brinkmann, 2022). Complementing these findings, survey-based evidence shows that blockchain and information governance strategies mediate security and trust in digital democratic processes, underscoring the importance of clear and comprehensive governance frameworks for applications such as electronic voting (Olaniyi, 2024). Perspectives from the developer community further extend this discussion, as interviews and focus groups with highly active GitHub contributors highlight the perceived need for blockchain governance across multiple public service domains, and emphasize transparency, immutability, and irreversibility as core governance principles (Kassen, 2024).

*4.3.3 Performance Metrics and Success Indicators*

Finally, empirical studies have examined blockchain performance and success following adoption in public sector contexts. For example, Dhingra et al. (2024) analyzed healthcare supply chain performance in India using survey data from 304 professionals across public and private healthcare organizations with experience in blockchain technologies. Using structural equation modeling, the study finds positive effects of blockchain adoption and implementation on healthcare supply chain performance. This evidence highlights the relevance of rigorous performance metrics and success indicators for assessing the impacts of blockchain use in the public sector.

Taken together, the findings across Sections 4.1 to 4.3 indicate that blockchain adoption in public services is embedded within polycentric governance arrangements characterized by multiple decision-making centers, overlapping institutional responsibilities, and context-dependent implementation choices. Descriptive patterns, implementation architectures, and empirical evidence collectively show that governments do not adopt blockchain as a uniform technological solution; rather, they selectively configure blockchain systems in response to sectoral demands, regulatory constraints, and institutional capacities. Across cases, authority, coordination, and accountability are distributed among public agencies and external actors, while core government institutions typically retain ultimate oversight and responsibility for public value outcomes. These observations suggest that the performance and governance of blockchain-enabled public services are shaped more by how governance roles and rules are structured across actors and levels of government and less by technical design alone. The following section builds on these findings to articulate their theoretical implications for polycentric governance and digital government.

## 5. Theoretical Implications for Polycentric Governance and Digital Government

Building on the empirical synthesis presented in Section 4, this study advances theory at the intersection of polycentric governance, digital government, and blockchain-enabled public services in three important ways. Rather than treating blockchain as a neutral or purely technical innovation, the findings demonstrate that blockchain adoption in national governments is embedded within, and actively reshapes, polycentric governance arrangements characterized by distributed authority, inter-organizational coordination, and layered accountability. By synthesizing evidence across public service domains and national contexts, this section articulates how blockchain-enabled systems reconfigure governance roles, rules, and responsibilities in digital government settings.

### 5.1 Reframing Blockchain as a Governance Infrastructure in Polycentric Systems

The first theoretical contribution lies in reframing blockchain from a technical infrastructure to a governance infrastructure within polycentric public service systems. Prior research on digital government has often examined blockchain in terms of efficiency gains, transparency enhancements, or security improvements. In contrast, this systematic review finds that blockchain architectures embed governance choices by encoding rules for coordination, validation, and information access directly into system design. From a polycentric governance perspective, blockchain-enabled public services function as institutional arrangements that shape how authority is exercised and coordinated across multiple decision-making centers. Smart contracts, consensus mechanisms, and permissioning structures formalize inter-organizational interactions that were previously governed through administrative procedures or informal coordination. This extends polycentric governance theory by demonstrating how digital infrastructures can operationalize governance principles and, in practice, constrain or enable institutional behavior. In doing so, the study contributes to digital government theory by highlighting the governance effects of technology design choices, rather than treating technology as an exogenous input.

### 5.2 Controlled Polycentricity and Configurational Variation in Digital Government

A second contribution concerns the identification of controlled polycentricity as a recurring governance pattern in blockchain-enabled public services. Classical polycentric governance theory emphasizes decentralization, autonomy, and experimentation among multiple centers of authority. The findings of this study refine this view by showing that, in national government contexts, decentralization is rarely absolute. Instead, governments deliberately configure blockchain systems to distribute execution and coordination while retaining centralized oversight and accountability. Across service domains, hybrid blockchain architectures, permissioned networks, and role-based access controls are used to balance decentralization with regulatory compliance and public value protection. These configurations vary according to sectoral sensitivity, legal constraints, and institutional capacity, resulting in multiple, context-dependent forms of polycentric governance rather than a single dominant model. This insight advances polycentric governance theory by moving beyond binary distinctions between centralized and decentralized systems and toward a configurational understanding of how digital government institutions manage authority and coordination under constraint.

### 5.3 Governance Tensions, Accountability, and Public Value in Blockchain-Enabled Services

A third theoretical implication relates to the governance tensions that emerge when blockchain technologies are embedded in polycentric public service systems. While blockchain can enhance transparency, traceability, and coordination, it also introduces rigidity through immutability and automated rule enforcement. The findings in this study show that accountability in blockchain-enabled public services increasingly shifts from ex post administrative control to ex ante governance through system design.

This shift creates tensions between algorithmic enforcement and administrative discretion, as well as between transparency and privacy in sensitive service domains. Polycentric governance theory has traditionally emphasized flexibility and adaptive learning; this study extends the theory by illustrating how digitally encoded governance rules can both support and constrain such adaptability. For digital government scholarship, these findings underscore that public value creation in blockchain-enabled services depends not only on technical performance but also on how accountability, oversight, and discretion are institutionally structured across actors and levels of government.

*5.4 Implications for Digital Government Theory*

These contributions strengthen the theoretical foundations of digital government research by integrating polycentric governance theory with empirical insights on blockchain adoption. The present study demonstrates that blockchain-enabled public services are not merely technological innovations layered onto existing institutions, but governance arrangements that reshape how authority, coordination, and accountability are distributed in the public sector. By foregrounding governance design choices and institutional context, this research advances digital government theory toward a more nuanced understanding of how emerging technologies transform public administration in polycentric environments.

## 6. Challenges and Opportunities for Blockchain-Based Technologies in the Public Sector

Building on the polycentric governance dynamics identified in the preceding section, this section synthesizes key challenges and opportunities associated with blockchain adoption in public services. Drawing on our systematic analysis of the extant literature, the section identifies a range of challenges that continue to constrain the adoption and use of blockchain-based technologies in the public sector, including technological, organizational, and institutional impediments. It further examines strategies proposed in the literature to mitigate these challenges and highlights opportunities to enhance the role of blockchain technologies in public service delivery.

**6.1 Challenges in Blockchain Adoption and Use**

*Technological challenges*

*Low throughput* and *high storage and maintenance costs* are the most significant technological challenges in financial blockchain applications (Vanmathi et al., 2024). These challenges present opportunities for governments to learn from these difficulties and to design effective non-financial blockchain systems with robust security, improved scalability, efficient consensus algorithms, and automated smart contract auditing to enhance daily operations. There are also opportunities to integrate blockchain with other cutting-edge technologies, such as advanced edge computing using Internet of Things (IoT) devices, machine learning, and artificial intelligence.

Another challenge concerns *security and privacy*, including private key management, vulnerabilities in smart contracts, quantum computing, and interoperability. However, researchers have demonstrated that these issues can be mitigated through multiple mitigation strategies, including secure key management, improved consensus, smart contract verification, quantum-resistant cryptography, interoperability standardization, and cross-chain communication protocols (Kuznetsov, 2024). Additionally, the creation of government regulatory bodies and the enforcement of governance rules can mitigate security and privacy issues related to blockchain use in the public sector (Phadke, Medrano & Ustymenko, 2022).

*Organizational challenges*

The main obstacle to implementing blockchain is the *lack of knowledge and skills* among public servants. This makes it difficult to build a trust network by bringing together diverse stakeholders, including ministries, agencies, researchers, and national and local administrators (Batubara et al., 2022). To foster collaboration and enhance the impact of blockchain-based e-government, it is essential to clearly identify who is involved in the trust network and understand their interests. Nonetheless, this can be difficult if stakeholders have limited knowledge and skills.

In addition, with the introduction of blockchain, tasks performed by public servants are shifting from traditional bureaucratic practices toward an increased focus on governing, maintaining, and developing blockchain applications within specific public services. The goal of these organizational changes may be to enhance coordination and communication between civil servants and other actors involved in the co-production and provisioning of public services. However, the managerial changes and automation of certain bureaucratic functions may result in *internal resistance to blockchain adoption* and *reduction of jobs*, posing significant costs for civil servants (Cagigas et al., 2021).

*Institutional challenges*

In many countries, *existing laws and regulations do not adequately support blockchain adoption* and use. The introduction of the General Data Protection Regulation (GDPR) in Europe highlights tensions between citizens' rights as data subjects and the exercise of data ownership and privacy. The most frequently cited conflicts in blockchain systems include difficulties in exercising the right to be forgotten or the right to erasure, problems in identifying data processors and controllers, challenges with data minimization and purpose limitation, and the territorial scope of public sector blockchain (Belen-Saglam, 2023). These challenges underscore the need for stronger laws and regulatory compliance systems to enhance the effectiveness of blockchain use in the public sector.

Finally, *infrastructure challenges* continue to constrain blockchain adoption and use in public services. For example, governments in North America that have experimented with blockchain in areas such as payment efficiency, health records management, and supply chain face various infrastructure challenges, including regulating competition and enhancing interoperability among different platforms. For African countries exploring blockchain, the primary challenge is an underdeveloped technological infrastructure, alongside various institutional and regulatory barriers. Similarly, South American countries face problems related to infrastructure availability (Pineda, Jabba & Nieto-Bernal, 2024).

## 6.2 Strategies and Opportunities for Addressing Blockchain Adoption Challenges

Scholars have identified several blockchain-based *safeguards to address trust concerns* in digital public service environments. These include (1) implementing robust monitoring systems; (2) engaging stakeholders in technology necessity discussions, (3) establishing knowledge repositories, (4) adopting mature and independent impact assessment approaches that respect public values and security, (5) conducting iterative, technology-agnostic evaluations, and (6) maintaining institutional expertise to prevent competence outsourcing (Bodo & Janssen, 2022).

Blockchain can bring positive changes to government culture through a higher form of accountability, allowing the public to monitor network activity. It can also reduce bureaucracy and paperwork in government administrative processes. One way to further enhance the contribution of blockchain in public service delivery is to *develop key performance indicators*. Inter-agency coordination and communication among civil servants and other key players using blockchain can be tracked by measuring the time spent on coordination activities and processes. In addition, governments must develop indicators to gauge the benefits of blockchain through its contribution to public value (Cagigas et al. 2023).

Additionally, *government support for public-private partnerships* (PPPs) is crucial to the successful implementation of blockchain across diverse domains in the public sector. For example, in land administration, blockchain-based applications seem to work best within a functioning governmental system that supports PPPs through requisite institutional changes and process redesign as well as regulatory frameworks (Saari, Vimpari & Junnila, 2022). Furthermore, governments' cybersecurity investments, alongside digital infrastructure and software development, are required to implement blockchain within the public procurement system, which also encourages private sector participation. The distinguishing features of public and private blockchains play an enormous role in the implementation of blockchains, where several access levels, depending on the stakeholders of the underlying process, should be addressed appropriately (Doguchaeva, Zubkova & Katrashova, 2022; Benchis, Shahzad & Dan, 2025).

## 7. Summary, Conclusions, and Recommendations for Governments

This systematic review provides a comprehensive synthesis of recent theoretical and empirical research on blockchain adoption in the public sector. Drawing on peer-reviewed studies published between 2021 and 2025 and analyzed in accordance with PRISMA guidelines, the review examines the current landscape of blockchain-based public service applications, along with their associated strengths, challenges, and opportunities. The findings suggest that blockchain adoption in public services unfolds within a polycentric governance environment characterized by multiple decision-making centers, overlapping institutional responsibilities, and diverse technological and organizational configurations. Against this background, this section summarizes key conclusions from the review and presents actionable recommendations for governments considering or advancing blockchain-based public service initiatives.

Our systematic review found that many countries are currently using or exploring the use of state-of-the-art decentralized applications based on blockchain technology for government service delivery. Accordingly, in Section 4.2, we presented a unified public service architecture that integrates prominent state-of-the-art blockchain platforms, such as Ethereum, Hyperledger Fabric, IPFS, and Corda, while incorporating complementary emerging technologies, including IoT, AI, and big data analytics. This architectural framework addresses critical interoperability and scalability challenges that have historically limited blockchain adoption in complex government environments. It also leverages the specific strengths of blockchain-based systems, including improved transparency and reduced administrative workload.

Furthermore, the present study found that in many countries, persistent challenges continue to affect the full potential of blockchain innovations in the public sector. These challenges pertain to scalability, technical complexity, underdeveloped technological infrastructure, regulatory compliance issues, lack of expertise among civil servants, and stakeholder resistance to organizational change. These barriers are especially dominant in developing countries.

As national governments worldwide consider blockchain innovations for public service delivery, this study identified critical factors that support the adoption and implementation of blockchain technologies across different domains of the public sector. These factors include government support, public-private partnership arrangements, existing digital identity solutions, societal involvement, comprehensive stakeholder engagement, adequate skills development programs, and regulatory frameworks that support innovation while ensuring appropriate oversight and compliance with data protection requirements. In addition, the evidence supports hybrid blockchain architectures that balance transparency requirements with privacy protection needs, implemented through phased approaches that allow for gradual capacity building and risk mitigation.

Despite growing empirical interest, several important research gaps remain. Future studies would benefit from longitudinal designs that examine blockchain implementation outcomes over extended periods, complementing existing work that emphasizes technical feasibility. The literature remains geographically concentrated in North America, Europe, and Asia, with limited empirical evidence from Africa, South America, and Oceania. Further research is also needed on blockchain applications in environmental and sustainability domains, particularly in climate governance and resource management. Finally, developing standardized evaluation frameworks would enable more systematic comparison of blockchain performance across sectors and institutional settings.

### *Recommendations for national governments*

We offer actionable guidance for government leaders and technology managers responsible for developing and executing blockchain initiatives. Based on the empirical evidence discussed in this article, we offer five key recommendations that address the need for (1) robust decision frameworks, (2) comprehensive best practice guidelines, (3) practical risk assessment and mitigation strategies, (4) governance and regulatory frameworks at the national level, and (5) international cooperation arrangements.

#### *(1) Establish robust decision frameworks for blockchain adoption*

Governments should establish robust, context-sensitive decision frameworks for blockchain adoption that assess both technical feasibility and organizational readiness. These frameworks should begin with a clear articulation of the specific problem to be addressed and evaluate whether blockchain is an appropriate solution relative to available alternatives. They should also include systematic stakeholder analysis and iterative consultation processes, ensuring that the perspectives of affected agencies, service users, and private-sector partners inform adoption decisions throughout the implementation lifecycle.

*(2) Develop comprehensive best practice guidelines*

Evidence from the literature indicates that governments should develop comprehensive yet adaptable best-practice guidelines for blockchain implementation. Hybrid blockchain architectures that balance transparency with privacy protection appear particularly well suited to public sector contexts. Phased implementation strategies and modular system designs enable agencies to build capacity incrementally, reduce implementation risks, and integrate blockchain applications with existing information systems. These guidelines should be supported by sustained training and capacity-building programs to strengthen long-term institutional capability.

*(3) Integrate practical risk assessment and mitigation strategies*

Effective blockchain adoption requires governments to integrate continuous risk assessment and mitigation into implementation processes. Technical risks related to scalability, security, and interoperability require ongoing monitoring, while organizational risks – such as skills gaps and resistance to change – necessitate active change management and workforce development. Regulatory and compliance risks should be addressed through adaptive governance arrangements that enable blockchain initiatives to evolve in response to changing legal and policy environments.

*(4) Adopt governance and regulatory frameworks*

Sustainable blockchain adoption depends on governance and regulatory frameworks that clearly define roles, responsibilities, and coordination mechanisms among public actors. As blockchain applications expand across multiple domains, such frameworks are essential to foster interoperability, accountability, and collaboration while preserving institutional autonomy. Effective governance arrangements can also support cross-sector digital platforms and productive partnerships with private-sector actors.

*(5) Strengthen international cooperation arrangements*

Given the cross-border nature of blockchain technologies, governments should strengthen international cooperation on blockchain governance. International coordination is critical not only for addressing illicit financial activity associated with blockchain-based assets but also for harmonizing regulatory standards, sharing technical expertise, and supporting interoperable governance frameworks. Such cooperation can enhance oversight capacity and enable governments to leverage blockchain technologies to advance transparency, accountability, and open governance objectives.

Overall, the findings of this systematic review underscore that effective blockchain adoption in the public sector depends not only on technological capability but also on governance arrangements that support coordination, adaptability, and institutional diversity. Viewed through a polycentric governance framework, blockchain-enabled public services can be approached as evolving systems of interdependent actors rather than centrally engineered solutions. Future research that examines long-term implementation outcomes, expands geographic coverage, and develops standardized evaluation frameworks will be essential for advancing both scholarly understanding and practical application of blockchain technologies in public governance.